\documentclass[twocolumn]{aastex631}
\usepackage[utf8]{inputenc}

\usepackage[version=4]{mhchem}
\usepackage{booktabs}
\usepackage{comment}
\usepackage{tabularx}

\usepackage{siunitx}
\usepackage{xcolor}
\usepackage{amsmath}
\usepackage[flushleft]{threeparttable}

\makeatletter
\def\blfootnote{\xdef\@thefnmark{}\@footnotetext}
\makeatother

\DeclareSIUnit\angstrom{\text {Å}}

\begin{document}

\title{Modeling the Effect of C/O Ratio on Complex Carbon Chemistry in Cold Molecular Clouds}

\author[0000-0002-4593-518X]{Alex N. Byrne}
\affiliation{Department of Chemistry, Massachusetts Institute of Technology, Cambridge, MA 02139, USA}

\author[0000-0002-5171-7568]{Christopher N. Shingledecker}
\affiliation{Department of Chemistry, Virginia Military Institute, Lexington, VA 24450, USA}

\author[0000-0003-4179-6394]{Edwin A. Bergin}
\affiliation{Department of Astronomy, University of Michigan, Ann Arbor, MI 48109, USA}

\author[0009-0008-1171-278X]{Martin S. Holdren}
\affiliation{Department of Chemistry, Massachusetts Institute of Technology, Cambridge, MA 02139, USA}

\author[0000-0002-0332-2641]{Gabi Wenzel}
\affiliation{Department of Chemistry, Massachusetts Institute of Technology, Cambridge, MA 02139, USA}
\affiliation{Center for Astrophysics \textbar{} Harvard \& Smithsonian, Cambridge, MA 02138, USA}

\author[0000-0003-2760-2119]{Ci Xue}
\affiliation{Department of Chemistry, Massachusetts Institute of Technology, Cambridge, MA 02139, USA}

\author[0000-0001-7111-0176]{Troy Van Voorhis}
\affiliation{Department of Chemistry, Massachusetts Institute of Technology, Cambridge, MA 02139, USA}

\author[0000-0003-1254-4817]{Brett A. McGuire}
\affiliation{Department of Chemistry, Massachusetts Institute of Technology, Cambridge, MA 02139, USA}
\affiliation{National Radio Astronomy Observatory, Charlottesville, VA 22903, USA}

\correspondingauthor{Alex N. Byrne, Brett A. McGuire}
\email{lxbyrne@mit.edu, brettmc@mit.edu}

\begin{abstract}
Elemental abundances, which are often depleted with respect to the solar values, are important input parameters for kinetic models of interstellar chemistry. In particular, the amount of carbon relative to oxygen is known to have a strong effect on modeled abundances of many species. While previous studies have focused on comparison of modeled and observed abundances to constrain the C/O ratio, the effects of this parameter on the underlying chemistry have not been well-studied. We investigated the role of the C/O ratio on dark cloud chemistry using the \texttt{NAUTILUS} code and machine learning techniques for molecular representation. We find that modeled abundances are quite sensitive to the C/O ratio, especially for carbon-rich species such as carbon chains and polycyclic aromatic hydrocarbons (PAHs). CO and simple ice-phase species are found to be major carbon reservoirs under both oxygen-poor and oxygen-rich conditions. The appearance of \ce{C3H4} isomers as significant carbon reservoirs, even under oxygen-rich conditions, indicates the efficiency of gas-phase \ce{C3} formation followed by adsorption and grain-surface hydrogenation. Our model is not able to reproduce the observed, gas-phase C/H ratio of TMC-1 CP at the time of best fit with any C/O ratio between 0.1 and 3, suggesting that the modeled freeze-out of carbon-bearing molecules may be too rapid. Future investigations are needed to understand the reactivity of major carbon reservoirs and their conversion to complex organic molecules.
\end{abstract}

\section{Introduction}
Within the context of interstellar chemistry, carbon plays a central role. Carbon comprises the backbone of all organic molecules, many of which are key ingredients for life on Earth. Although a sizable portion of interstellar carbon is thought to exist in atomic/ionic form or as CO depending on the visual extinction of the region, 10-25\% is thought to be contained in polycyclic aromatic hydrocarbons (PAHs) \citep{dwek_detection_1997, habart_pahs_2004, tielens_interstellar_2008, chabot_coulomb_2019}. Additionally, the detections of numerous complex organic molecules within the interstellar medium (ISM) indicate that some portion of carbon is contained within these species \citep{mcguire_2021_2022, yang_corinos_2022, mcclure_ice_2023, salyk_corinos_2024, xue_molecular_2025}. Many of these complex organic molecules have been detected at the cyanopolyyne peak in the cold and dense Taurus Molecular Cloud 1 (TMC-1 CP), evidencing that complex carbon chemistry can occur even at very low temperatures. In particular, CN-derivatives of naphthalene \citep{mcguire_detection_2021}, acenaphthylene \citep{cernicharo_discovery_2024}, pyrene \citep{wenzel_detection_2024, wenzel_detections_2024} and coronene \citep{wenzel_discovery_2025} have recently been detected in TMC-1 CP. The pure hydrocarbons pyrene and coronene are estimated to each contain approximately 0.04\% of the total carbon when accounting for depletion onto dust grains, suggesting that individual PAHs can be significant reservoirs of carbon within cold molecular clouds. This would be a significant component of the interstellar PAH population if this material is returned to the ISM. However, kinetic models built to simulate the chemical evolution of dense clouds have historically failed at reproducing the observed abundances of carbon-containing molecules, particularly larger ones \citep{byrne_astrochemical_2023}.

When investigating the interstellar formation of carbon-bearing molecules, it is first necessary to have an understanding of the elemental abundances available within that region. Observations of atomic absorption lines toward various regions of the ISM have found that gas-phase atomic abundances of carbon and heavier elements are depleted with respect to the solar abundances - the elemental abundance ratios in our solar system based on measurements of the Sun's photosphere and carbonaceous meteorites \citep{morton_spectrophotometric_1973, morton_interstellar_1974, lodders_solar_2003, jenkins_unified_2009}. This depletion is thought to relate to the presence of these elements as key components of refractory solids or dust grains \citep{morton_interstellar_1974}. Using gas-phase elemental abundances observed toward 243 sight lines, \citep{jenkins_unified_2009} derived a relationship between the depletion of a particular element and the overall depletion of a sight line. These depletion factors serve as a baseline for estimating gas-phase elemental abundances of interstellar sources.

Historically, kinetic models of dense clouds have often adopted elemental abundances based on those of the diffuse cloud $\zeta$ Oph as it is a well-studied source with observed depletion in many elements \citep{graedel_kinetic_1982, savage_ultraviolet_1992, jenkins_unified_2009}. Updated values from the analysis by \citet{jenkins_unified_2009}, preferred by many recent studies, result in a C/O ratio of $\sim0.55$ \citep{hincelin_oxygen_2011, agundez_chemistry_2013}. It has long been known that kinetic models of interstellar chemistry are highly sensitive to this parameter, including isotopic fractionation \citep{langer_carbon_1984, langer_ion-molecule_1989} as well as the abundances of many carbon-containing species, from atomic carbon to complex organic molecules \citep{herbst_ion-molecule_1983,langer_carbon_1984, herbst_synthesis_1986, langer_ion-molecule_1989, wakelam_sensitivity_2010, hincelin_oxygen_2011, loomis_investigation_2021}. Although the solar abundance of oxygen is greater than that of carbon by a factor of two \citep{lodders_solar_2003}, multiple mechanisms have been proposed that could modify the gas-phase C/O ratio as a diffuse cloud collapses to a dense cloud. For example, the efficient freeze-out of water onto interstellar grains at low temperatures could deplete oxygen from the gas-phase as the cloud collapses \citep{furuya_water_2015, ruaud_influence_2018}. In conjunction with a some mechanism of supplying carbon, such as destruction of carbonaceous grains \citep{bocchio_re-evaluation_2014} or CO \citep{sellek_chemical_2025}, this could lead to enhanced C/O ratios above one at the beginning of the dense cloud phase.

It has been found that a larger C/O ratio than originally used often provides better agreement between kinetic models and observations \citep{herbst_ion-molecule_1983, watt_observations_1986, swade_physics_1989, bergin_chemical_1995}. \citet{hincelin_oxygen_2011} studied the effects of a decreasing initial oxygen abundance on a kinetic model of dark cloud chemistry. The authors found that an additional depletion of over a factor of 2 in oxygen was necessary to produce modeled \ce{O2} abundances in agreement with observations. They also found that this depleted oxygen abundance resulted in better overall agreement between modeled and TMC-1 CP observed abundances, partially due to larger abundances of cyanopolyynes (\ce{HC$_n$N}, $n =$ odd). Likewise, \citet{loomis_investigation_2021} performed kinetic modeling of cyanopolyynes up to \ce{HC11N} in a dark cloud using a range of oxygen abundances and thus C/O ratios, finding that a ratio of 1.1 provides excellent agreement with TMC-1 CP observations whereas a ratio of 0.7 results in considerably poor agreement. \citet{agundez_chemistry_2013} compared the fraction of reproduced molecules for 70 TMC-1 CP abundances using the 2011 KIDA and UMIST RATE12 chemical networks and C/O ratios of 0.55 and 1.4. Contrary to the aforementioned studies, the authors found that the C/O ratio of 0.55 resulted in a greater percent of molecules reproduced by the model, although only these two ratios were considered. Furthermore, the chemical networks used have undergone substantial changes since such as the inclusion of additional complex organic molecules, refinement of reactions and rate coefficients, and incorporation of ice-phase chemistry.

It thus remains unclear which elemental abundances of carbon and oxygen best represent the amounts present in TMC-1 CP, with  proposed C/O ratios spanning a large range from 0.55 to 1.4 \citep{agundez_chemistry_2013}. Furthermore, while previous analyses have focused on the effect of C/O ratio on agreement with observations, there has been less emphasis on the influence that this parameter has on the underlying chemistry. Such an understanding is vital as the number of detected molecules, including long carbon chains and PAHs, continues to grow. A comprehension of the chemical effects of the C/O ratio is also necessary for generalizing results from TMC-1 CP to other dark clouds or to the following protoplanetary disk stage. 

Here we present an analysis on the influence of the C/O ratio on dark cloud chemistry using kinetic modeling and machine learning. In Section~\ref{sec:Methods} we describe the modeling code and parameters used, the modification of the C/O ratio, and the implementation of machine learning embedding to visualize the molecules in the network. In Section~\ref{sec:Results} we show the sensitivities of carbon-bearing molecules to the C/O ratio, the time-dependent carbon fractions of major carbon reservoirs, and the partitioning of carbon between gas and ice phase as a function of C/O ratio. In Section~\ref{sec:Discussion} we discuss the implications of these results for interstellar carbon chemistry. Finally, in Section~\ref{sec:Conclusions} we summarize the major findings and propose areas for further investigation.

\section{Methods}
\label{sec:Methods}
We use the 2016 version of the \texttt{NAUTILUS} 3-phase modeling code, an open-source rate equation code \citep{ruaud_gas_2016}. This code has been extensively used in modeling of molecular clouds such as TMC-1 CP \citep[e.g.][]{2018ApJ...861...20S, 2020A&A...637A..39N}. \texttt{NAUTILUS} treats the gas phase, grain surface, and grain mantle as separate phases, each with their own set of chemical reactions. Rates of adsorption onto grain surfaces, desorption from grain surfaces, and swapping between grain surface and mantle are also computed. The considered desorption mechanisms consist of thermal evaporation, evaporation via cosmic-ray heating, photodesorption by external and internal, cosmic-ray induced photons, and chemical desorption due to reaction exothermicity \citep{garrod_non-thermal_2007}. The physical conditions used are those considered typical of a cold dark cloud, consisting of a kinetic temperature of 10\,K for the gas and dust grains \citep{pratap_study_1997, feher_structure_2016, fuente_gas_2019}, a gas number density of \SI{2e4}{\per\cubic\centi\metre} \citep{snell_determination_1982, feher_structure_2016, fuente_gas_2019}, a cosmic-ray ionization rate of \SI{1.3e-17}{\per\second} \citep{spitzer_heating_1968, dalgarno_galactic_2006, padovani_cosmic-ray_2009}, and a visual extinction of 10\,mag for external UV photons \cite{fuente_gas_2019, rodriguez-baras_gas_2021}. The initial elemental abundances chosen are those from \citet{byrne_astrochemical_2023}. These are primarily the ``low-metal'' abundances from \citet{graedel_kinetic_1982}. The gas- and ice-phase reactions, hereafter referred to as chemical networks, are those used and described in \citet{byrne_sensitivity_2024}. These networks are based on kida.uva.2014 \citep{wakelam_kinetic_2012} and \citet{ruaud_modelling_2015} respectively and include a number of changes related to recently-detected species such as large carbon chains and aromatic molecules. For example, our network considers formation of cyanonaphthalene (\ce{C10H7CN}) through reactions involving benzene (\ce{C6H6}) and phenyl radical (\ce{C6H5}), along with multiple pathways to forming six-membered aromatic rings.

To assess the effect of the C/O ratio on model results, 100 equally-spaced values for the C/O ratio over a range of 0.1 to 3.0 were tested as functions of varying initial carbon and oxygen abundances. These variations to the initial C/O ratio can be considered as a simple proxy for some mechanism that modifies the gas-phase elemental abundances as the dense cloud forms, for which a more detailed consideration is beyond the scope of this work. For every C/O ratio, two iterations of the model were performed: one where the ratio was achieved by increasing or decreasing the initial carbon abundance from its nominal value of $1.70\times10^{-4}$, and one where the ratio was achieved by increasing or decreasing the initial oxygen abundance from its nominal value of $1.55\times10^{-4}$. Our nominal oxygen abundance is lower than the $\zeta$ Ophiuchi value of \citet{jenkins_unified_2009} as it has been found to better reproduce the TMC-1 CP abundances of cyanopolyynes \citep{loomis_investigation_2021}. For example, to test a C/O ratio of 0.5, we create two models: one with initial carbon and oxygen abundances of $7.75\times10^{-5}$ and $1.55\times10^{-4}$ respectively, and another with initial carbon and oxygen abundances of $1.70\times10^{-4}$ and $3.40\times10^{-4}$ respectively. In this way the initial abundances of carbon and oxygen were never changed concurrently. Time-dependent sensitivities of modeled abundances to C/O ratio were determined by calculating the relative standard deviation of these abundances as a function of C/O ratio at select times. This same metric was found to be a valid measure of sensitivity of modeled abundances to reaction rate coefficients \citep{byrne_sensitivity_2024}. It is widely believed that the ``chemical age'' of TMC-1 CP lies between $10^5$ and $10^6$ years based on agreement between kinetic models and observations \citep{hincelin_oxygen_2011, agundez_chemistry_2013, ruaud_gas_2016}. Recent modeling studies find the best agreement at a time of $\sim5\times10^5$ years, especially for more complex molecules such as PAHs and cyanopolyynes \citep{loomis_investigation_2021, chen_chemical_2022, byrne_astrochemical_2023}. For the proceeding analyses we thus focus on a time of 505,500 years, hereafter referred to as $5\times10^5$ years.

In order to better visualize and analyze the resulting data, we made use of machine learning and data science techniques. We used a pipeline similar to that used in prior machine learning of interstellar chemical inventories and molecular property prediction \citep{lee_language_2021, pfried_implementation_2023, scolati_explaining_2023, toru_shay_exploring_2025, marimuthu_machine_2025}. First, a SMILES (Simplified Molecular Input Line Entry System) string is supplied for each carbon-containing species in our chemical network. These strings are a convenient, machine-readable format for writing out chemical structures \citep{weininger_smiles_1988}. The SMILES strings are then converted to SELFIES strings, which are a similar representation but more robust and designed specifically for use in machine learning \citep{krenn_self-referencing_2020}. Next, the SELFIES strings are passed through the VICGAE embedder where each SELFIES string is converted to a 32-dimensional vector. This is a pretrained, astrochemistry-focused embedding model that captures chemical characteristics of each molecule within the dense, 32-dimensional vectors created. Information regarding properties such as size, connectivity, charge, and degree of saturation is encoded into these vectors, although individual vector dimensions themselves do not map directly to properties. This treatment is particularly useful for comparing the ``chemical similarity'' of multiple species, as the distance in vector space between two molecules corresponds to their similarity as learned by the VICGAE embedder. In order to visualize this vector space in two dimensions, we finally make use of the UMAP (Uniform Manifold Approximation and Projection; \citet{sainburg_parametric_2021}) algorithm to determine a set of two-dimensional vectors that best matches the structure of the 32-dimensional vector space. We use 75 neighbors, the euclidean metric, a minimum distance of 0.1, and a spread of 1 as our UMAP parameters. The end result is a set of two-dimensional vectors for each carbon-containing species that can be plotted to visualize the entire chemical network at once in terms of machine-learned chemical similarity.

\section{Results}
\label{sec:Results}
\subsection{Model sensitivity to C/O ratio}
\label{sec:sens}

\begin{figure*}
    \centering
    \includegraphics[width=\textwidth]{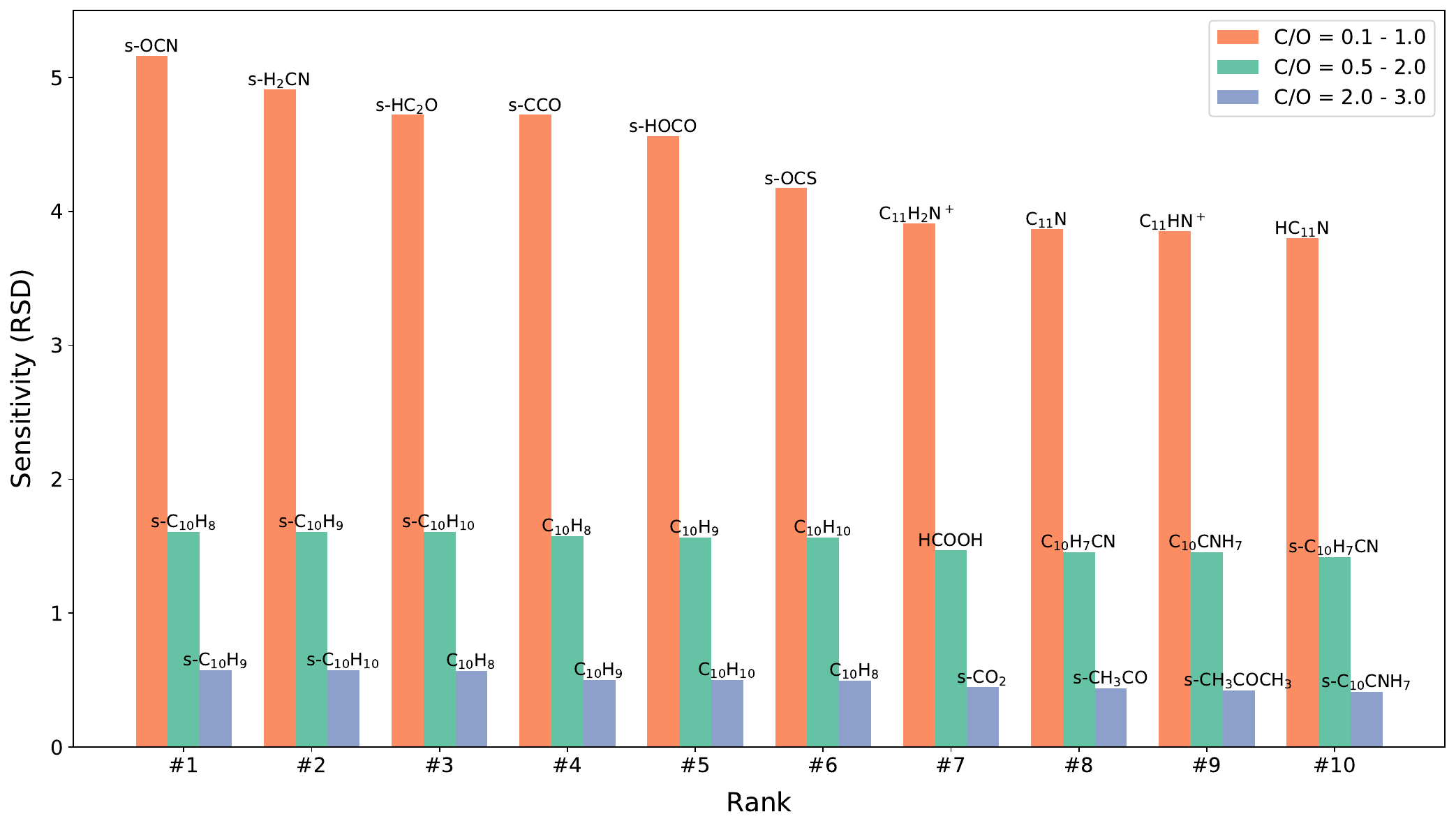}\hfill
    \caption{Top 10 most sensitive carbon-containing species to the C/O ratio at $5\times10^5$ years according to relative standard deviation. Relative standard deviations are shown for three different ranges of C/O ratio as discussed in the text: $\rm{C/O} = 0.1 - 1.0$ (orange), $\rm{C/O} = 0.5 - 2.0$ (green), and $\rm{C/O} = 2.0 - 3.0$ (blue). Each bar is annotated with the formula of the corresponding species as it appears in the model. Species names with a prefix of ``s-'' correspond to ice-phase species where the abundances in the surface and mantle phases have been summed together}, while the absence of a prefix indicates a gas-phase species.
    \label{fig:C_O_Sens}
\end{figure*}

For each species, the modeled abundances of that species at $5\times10^5$ years were collected from all models with varying C/O ratios. The relative standard deviation was calculated for each species by determining the standard deviation of these abundances and then dividing by the average. This relative standard deviation metric is a normalized measure of the spread in abundances as a function of C/O ratio. A larger relative standard deviation thus indicates that modeled abundances change more drastically as a function of C/O ratio, leading to a greater spread in the data. In Figure~\ref{fig:C_O_Sens} the 10 most sensitive species to the C/O ratio at $5\times10^5$ years according to this metric are shown for three different ranges of C/O ratios. This data is from the models where only the initial oxygen abundance has been changed, as this is the manner in which the C/O ratio is typically changed. Changing the initial carbon abundance can lead to additional effects as both the amount of carbon relative to oxygen and absolute amount of carbon are changing \citep{kanwar_minds_2025}. These regimes consist of C/O ranges of 0.1 to 1.0, 0.5 to 2.0, and 2.0 to 3.0. Hereafter the terms ``oxygen-rich'' and ``oxygen-poor'' are used to refer generally to models where the C/O ratio is below or above unity respectively as a result of changing the initial oxygen abundance. We note that the range of $0.1 - 1.0$ encompasses the entire oxygen-rich range tested, while the range of $2.0 - 3.0$ is a specific subset of oxygen-poor conditions that is substantially above unity. This is because there is significantly different effects of changing C/O ratio on modeled abundance at C/O ratios well above unity compared to C/O ratios near unity. The range of $0.5 - 2.0$ can roughly be considered an ``astrophysically reasonable'' range as it encompasses the range of C/O ratios used in other studies of dense clouds, typically spanning from 0.5 to 1.4 \citep{agundez_chemistry_2013, chen_chemical_2022}. While C/O ratios approaching 2.0 are larger than those typically used for dense clouds, they are consistent with models and observations of protoplanetary disks \citep{bergin_hydrocarbon_2016, kanwar_minds_2024, sellek_chemical_2025}.

The C/O = $0.1 - 1.0$ regime is characterized by large relative standard deviations for two main classes of molecules. The first is small, oxygen-bearing molecules such as OCN and \ce{H2CN} in the ice phase. The abundances of these species quickly drop as the C/O ratio rises due to oxygen depletion. The second is long, unsaturated carbon-chain species with 9-11 carbon atoms in the gas phase. As the window of C/O ratios shifts toward greater values, the overall relative standard deviations decrease and the identities of the most sensitive species change. In the C/O = $0.5 - 2.0$ range, the greatest relative standard deviations are now $\sim1.6$ for the two-ring PAH naphthalene (\ce{C10H8}) and the hydronaphthalenes (\ce{C10H10}, \ce{C10H9}, both in the gas phase and in the ice. These bicyclic PAHs are the only PAHs currently included in the model; larger PAHs whose cyano derivatives have been detected, specifically pyrene, coronene, and acenaphthylene, have not yet been added. When the C/O ratio is increased above 2.0, these species remain the most affected although the relative standard deviations are much lower at $\sim0.5$. A few ice-phase, oxygen-bearing molecules, specifically \ce{CO2}, \ce{CH3CO}, \ce{CH3OCH3} also appear here. The sharp decrease to relative standard deviations as the C/O ratio increases shows that modeled abundances are less sensitive to a changing oxygen abundance as oxygen becomes more depleted from the gas phase. It is under oxygen-rich conditions where changing the initial oxygen abundance has the strongest impact on modeled abundances. Recently we performed a sensitivity analysis of an astrochemical model to the rate coefficients of gas-phase reactions \citep{byrne_sensitivity_2024}. We found that the greatest sensitivity of a species abundance to an individual rate coefficient was a relative standard deviation of 1.43. Overall, abundances in our model are substantially more sensitive to the C/O ratio than any individual rate coefficient, especially within the regime of $0.1 - 1.0$.

\begin{figure}
    \centering
    \includegraphics[width=\columnwidth]{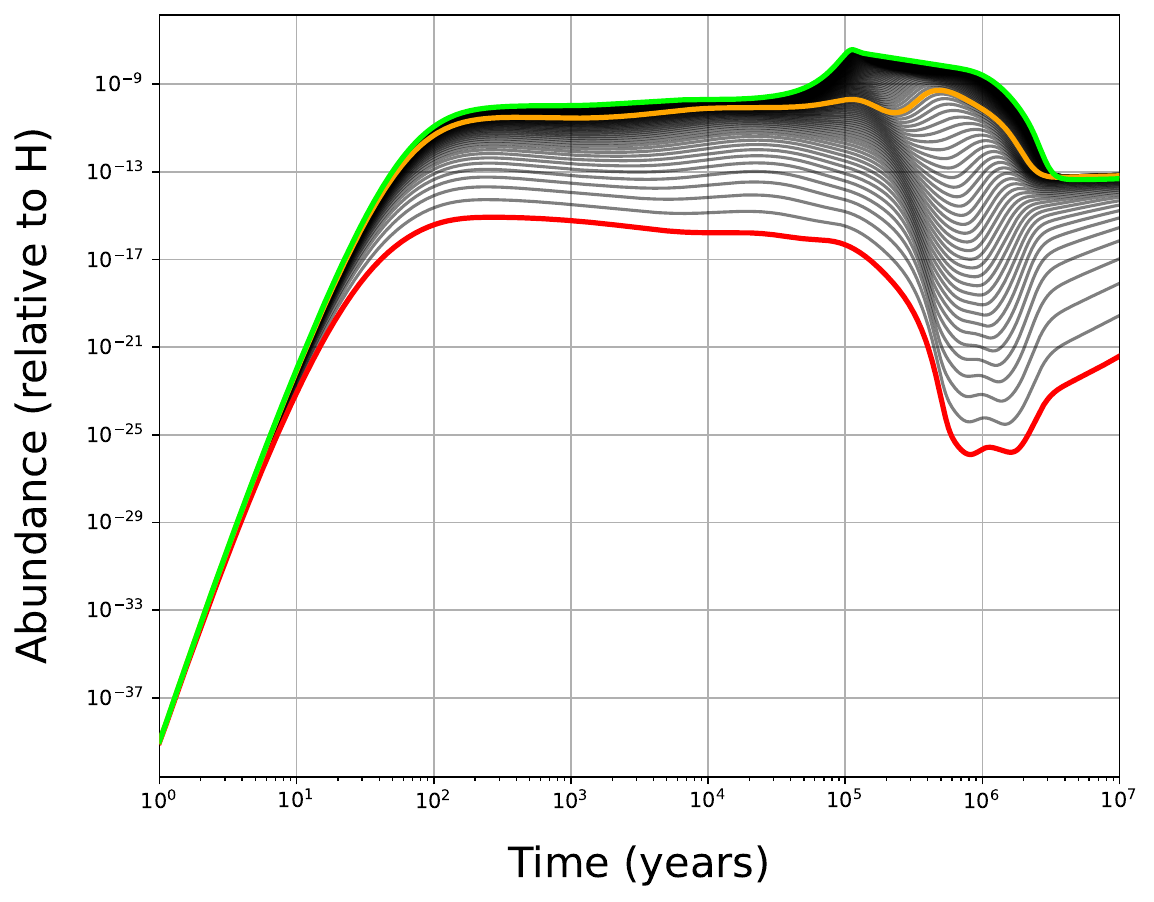}
    \caption{Modeled abundances relative to H of \ce{C11} as a function of time for every tested C/O ratio. The minimum ratio of 0.1 is shown in red, the nominal ratio of 1.1 is shown in orange, and the maximum ratio of 3.0 is shown in green.}
    \label{fig:C11}
\end{figure}

\begin{figure}
    \centering
    \includegraphics[width=\columnwidth]{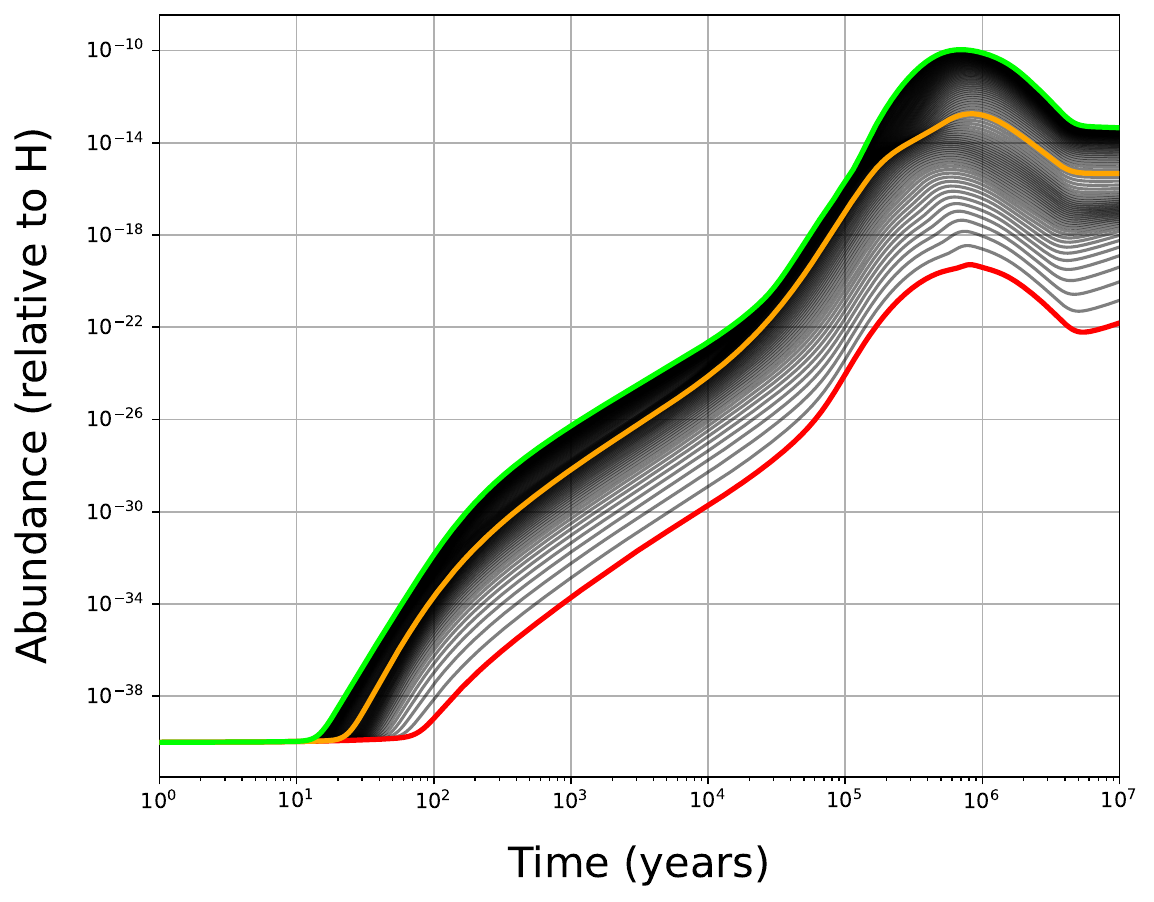}
    \caption{Same as Figure~\ref{fig:C11} but for \ce{C10H8}.}
    \label{fig:C10H8}
\end{figure}

The modeled abundances with respect to hydrogen nuclei of \ce{C11} and \ce{C10H8} as a function of time for all tested C/O ratios are shown in Figures~\ref{fig:C11} and \ref{fig:C10H8}, again with models where only the initial oxygen abundance was modified. In addition to the minimum and maximum C/O ratios tested, the abundance profile resulting from our ``nominal'' C/O ratio of 1.1 is shown in orange. This value, obtained by lowering the initial oxygen abundance to $1.55\times10^{-4}$, has been found by \citet{loomis_investigation_2021} to well reproduce the abundance of cyanopolyynes and is the typical value used in our model. Between $10^5$ and $10^6$ years the abundance of \ce{C11} drops substantially in models where the C/O ratio is low, with almost a 15 order of magnitude difference between the nominal and lowest abundances. Conversely, the abundance of \ce{C10H8} does not show a marked change in abundance profile but rather diminished production, with a $\sim$7 order of magnitude difference in nominal and minimum abundance in the aforementioned time range. Increasing the C/O ratio shows diminishing returns in terms of abundance increases in both species. For \ce{C11}, the abundance at $5\times10^5$ years is only about an order of magnitude larger in the model with a C/O ratio of 3 compared to the nominal model. Conversely, the abundance of \ce{C10H8} increases by a little over three orders of magnitude under the same circumstances. The diminishing changes to abundances, such as those of \ce{C11} and \ce{C10H8}, as the C/O ratio is increased up to 3.0 is why the relative standard deviations are significantly lower in the C/O = 2.0-3.0 regime.

\subsection{Carbon Reservoirs and carbon chemistry}

\begin{figure}
    \begin{interactive}{animation}{Figures/C_O_0.5.mp4}
    \includegraphics[width=0.9\columnwidth]{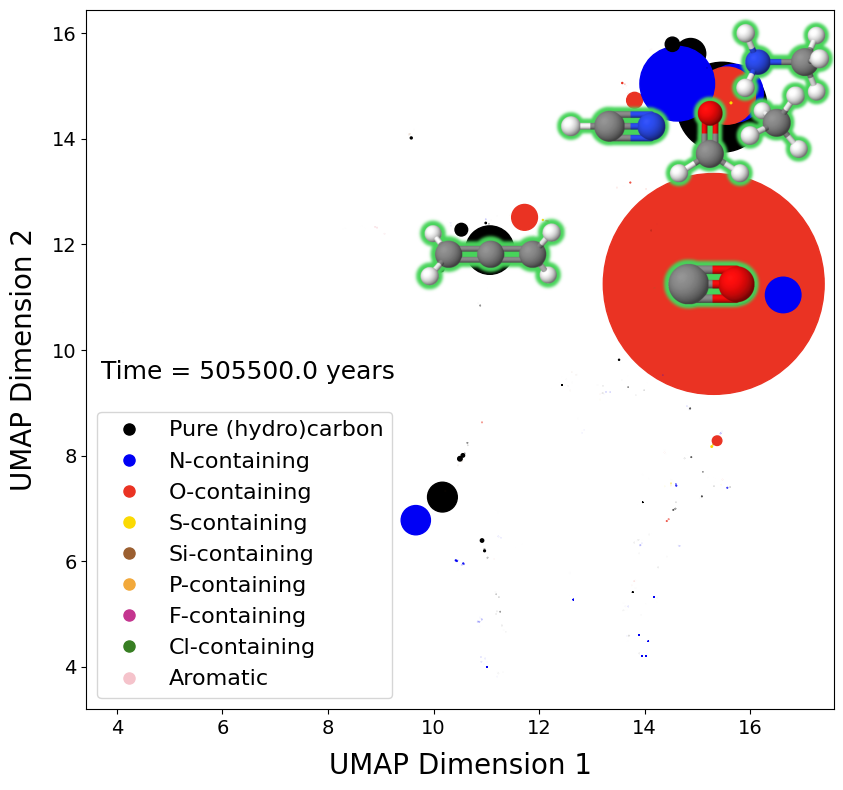}
    \end{interactive}
\caption{Modeled time-dependent carbon fractions of every carbon-containing species in the network with a C/O ratio of 0.5. Each dot represents a species that contains carbon plotted according to its UMAP vectors. The color corresponds to the elemental composition, while the size is the fraction of carbon multiplied by a scaling factor. Molecular structures are shown for species comprising at least 2\% of the total carbon. Some of structures are slightly offset from their corresponding dots for visual clarity. Structures for ice-phase (grain surface or mantle) species are highlighted in green. At the beginning of the simulation, nearly all of the carbon is in atomic form as represented by a large black dot in the center-right of the figure. Starting at $\sim100$ years, a red dot slightly to the left of atomic carbon corresponding to CO begins to grow. At $500$ years, a handful of dots clustered near the upper right corner also begin to slowly grow. These correspond to ice-phase \ce{HCN}, \ce{H2CO}, \ce{CH4}, and \ce{CH3NH2}. These species continue to grow in carbon fraction as time progresses, with CO reaching the 2\% of total carbon threshold at 4695 years. Starting at $10^4$ years, the dot corresponding to atomic carbon begins to rapidly diminish while the aforementioned dots grow more rapidly. A black dot below CO corresponding to \ce{C3} quickly grows from this point, reaching the 2\% threshold at 11220 years. At 19420 years ice-phase CO also reaches the 2\% threshold. At $3\times10^4$ years ice-phase \ce{H2CO} reaches 2\% of the total carbon budget, followed shortly by HCN, \ce{CH3NH2}, and \ce{CH4} all in the ice phase as well. At $10^5$ years atomic carbon begins to rapidly shrink in abundance, followed shortly by \ce{C3}. Gas-phase CO peaks in size at $\sim2\times10^5$ years and then decreases whereas ice-phase CO rapidly expands. There are also a few dots representing pure hydrocarbons, oxygen-containing species, and nitrogen-containing species that become sizable in terms of carbon fraction. The only one of these that reaches the 2\% threshold is ice-phase \ce{CH2CCH2} at $3.54\times10^5$, located a bit to the left of CO and growing in size until $\sim10^6$ years. Between $10^6$ years and the end of the simulation at $10^7$ years there are few changes except a decline in \ce{CH3NH2}, the appearance of ice-phase HNC as significant carbon reservoir near CO and atomic carbon, and continued growth of ice-phase \ce{CH4}. The static image displays the major reservoirs of carbon (those containing at least 2\% of the total carbon) at $\sim5\times10^5$ years, where our model best reproduces the abundances of larger carbon-bearing molecules.} 
    \label{fig:C-poor-Cfrac}
\end{figure}

\begin{figure}
    \begin{interactive}{animation}{Figures/C_O_1.0.mp4}
    \includegraphics[width=0.9\columnwidth]{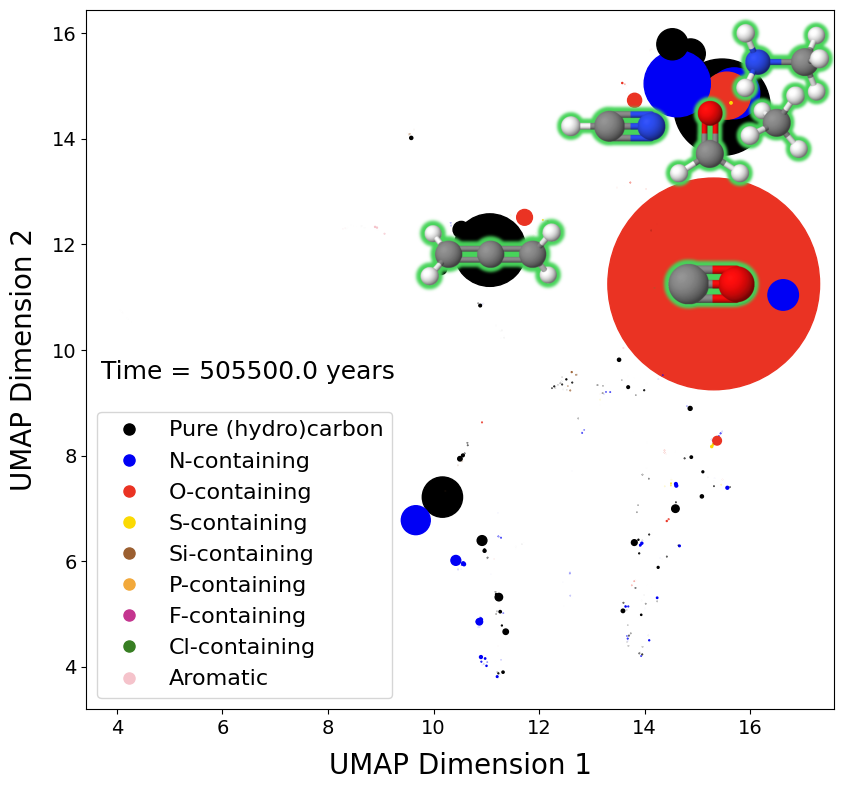}
    \end{interactive}
\caption{Same as Figure~\ref{fig:C-poor-Cfrac} but with a C/O ratio of 1. The evolution of carbon fractions begins very similarly to that of a C/O ratio of 0.5, although CO forms more slowly and does not reach the 2\% threshold until 5698 years. \ce{C3} reaches this same threshold immediately after at a time of 5791 years. As in Figure~\ref{fig:C-poor-Cfrac}, atomic carbon begins to quickly diminish at $10^4$ years whereas \ce{C3}, \ce{CO}, ice-phase CO, and ice-phase \ce{HCN}, \ce{H2CO}, \ce{CH4}, and \ce{CH3NH2} grow in size more quickly. Ice-phase CO achieves a carbon fraction of 2\% at 26390 years, followed by HCN, \ce{H2CO}, \ce{CH4}, and \ce{CH3NH2} at $\sim4\times10^4$ years. By $10^5$ years \ce{C3} is the second largest carbon reservoir next to CO, almost triple the size as in the model with a C/O of 0.5 at this time. Following this time there is rapid gain of ice-phase CO and loss of atomic carbon and \ce{C3} followed by gas-phase CO, although \ce{C3} maintains a carbon fraction of at least 2\% until a time of $3.78\times10^5$ years compared to $3.02\times10^5$ years in Figure~\ref{fig:C-poor-Cfrac}. A growth in various nitrogen-containing, oxygen-containing, and pure hydrocarbon species is again seen although at an earlier time and to a stronger extent. \ce{CH2CCH2} reaches the 2\% threshold at $1.58\times10^5$ years followed by its isomer \ce{CH3CCH} at $1.83\times10^5$ years. The dot for \ce{CH3CCH} is located a good distance below that of \ce{CH2CCH2} at the top of a series of dots that extends down to the bottom of the graph. Between $10^5$ and $\sim5\times10^5$ many dots in this series, along with another series on the right side of the figure below CO, grow in size sequentially, although none of them reach a carbon fraction of 2\%. The carbon fraction of \ce{CH3CCH} diminishes from $3\times10^3$ years and drops below 2\% by $4.37\times10^5$ years, while the fraction of carbon in \ce{CH2CCH2} continues to grow. Between $10^5$ years there is also a decline in ice-phase \ce{CH3NH2} and a growth in ice-phase \ce{CH4}. From $10^6$ to $10^7$ years the major observations are the appearance of ice-phase \ce{C3H8} as a major carbon reservoir at $10^6$ years in the top-right cluster, a continued growth in ice-phase \ce{CH4}, and the continued diminishing of various other species such as the series of dots that extend toward the bottom of the figure.}
    \label{fig:C-unity-Cfrac}
\end{figure}

\begin{figure}
    \begin{interactive}{animation}{Figures/C_O_2.0.mp4}
    \includegraphics[width=0.9\columnwidth]{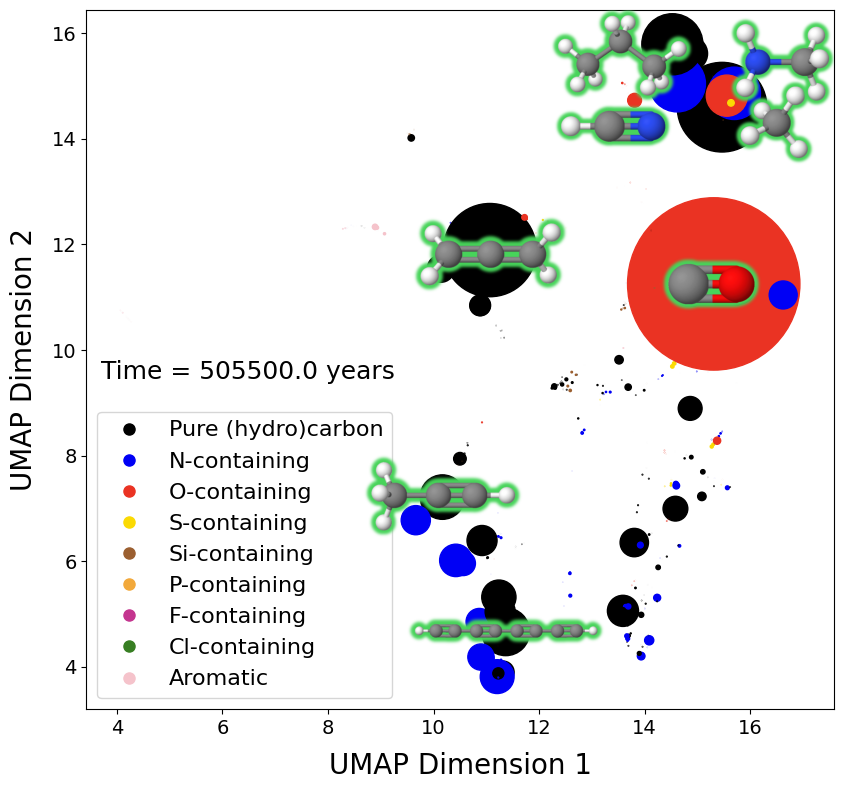}
    \end{interactive}
\caption{Same as Figures~\ref{fig:C-poor-Cfrac} and \ref{fig:C-unity-Cfrac} but with a C/O ratio of 2.0. Compared to the previous figures there is an even faster growth of \ce{C3} here, with it reaching the 2\% threshold at 3869 years even before CO at 5607 years. Between $10^4$ and $10^5$ years the carbon fraction of atomic carbon again decreases while the fractions in CO and \ce{C3} grow. The sizes of CO and \ce{C3} are almost equal over this range of time. Again ice-phase \ce{CH3NH2}, \ce{CH4}, \ce{H2CO}, and \ce{HCN} become major carbon reservoirs, although here \ce{CH3NH2} is the first to reach 2\% at $\sim3\times10^4$ years. This is followed shortly by \ce{CH4} at $\sim4\times10^4$ years, and then HCN and \ce{H2CO} at $\sim8\times10^4$ years. Over this range of $10^4$ to $10^5$ years there can also be seen the growth of nitrogen-containing and pure hydrocarbon species near \ce{CH2CCH2} and \ce{CH3CCH}. Following $10^5$ years the carbon fractions in gas-phase CO and \ce{C3} again diminish as ice-phase CO grows, however there are a number of differences compared to the previous figures. The sequential growth in the nitrogen-containing and pure hydrocarbon species below \ce{CH3CCH} and below CO is much more pronounced. At $3.32\times10^5$ years, one of these dots corresponding to \ce{C8H2} reaches 2\% of the total carbon, although a number of nearby dots appear to be only slightly smaller in size. These species again diminish in size following $5\times10^5$ years, except for ice-phase \ce{C8H3} which appears at $sim9\times10^5$ years and remains until the end of the simulation. In the top-right corner ice-phase \ce{H2CO} drops below 2\% of all carbon shortly after $10^5$ years but returns at $8\times10^5$ years, while ice-phase \ce{C3H8} reaches 2\% at $2.45\times10^5$ years. From $10^6$ until $10^7$ years, ice-phase \ce{CH2CCH2} and \ce{CH4} continue to grow in size. A few new species additionally appear as major carbon reservoirs. Ice-phase \ce{CH2CHCCH} appears near \ce{CH2CCH2} shortly after $10^6$ years, followed quickly by ice-phase \ce{C2H6} near \ce{CH4} and \ce{C3H8}. Finally, \ce{C6H4} located below CO becomes a major reservoir of carbon at $1.35\times10^6$ years.}
    \label{fig:C-rich-Cfrac}
\end{figure}

In Figures~\ref{fig:C-poor-Cfrac}, \ref{fig:C-unity-Cfrac}, and \ref{fig:C-rich-Cfrac} the modeled time-evolution of fraction of carbon in all carbon-containing species is shown for models with oxygen-rich conditions (C/O = 0.5), equal carbon and oxygen (C/O = 1), and oxygen-poor conditions (C/O = 2.0) respectively. As in Section~\ref{sec:sens}, we use the models where only initial oxygen abundance is changed to focus on C/O ratio as a function of oxygen depletion. These figures display every carbon-containing molecule in our network plotted as its UMAP dimensions obtained from the procedure described in Section~\ref{sec:Methods}. As mentioned previously, the values of these UMAP dimensions do not directly convey any molecular information. Instead, they represent a molecule's position in a machine-learned vector space based on properties such as size, and connectivity. In this way, a large number of molecules can be visualized concurrently with the distances between them proportional to their chemical similarities as learned by the VICGAE embedder, which may differ from human intuition. Each dot in these figures corresponds to one molecule, while the size of the dot is proportional to the fraction of total carbon contained in this molecule at the displayed time. The color of the dot corresponds to the elemental composition, namely the presence of heteroatoms or aromaticity. For molecules containing multiple heteroatoms or containing a heteroatom and aromaticity, aromaticity takes highest priority for the color coding, followed by the presence of silicon, sulfur, oxygen, and nitrogen. There are no carbon-containing species in the model that have a phosphorus, fluorine, or chlorine atom and are aromatic or contain another heteroatom.

Starting at approximately 1000 years, the primary form of carbon begins shifting from gas-phase atomic form to simple gas- and ice-phase species. Between $\sim5\times10^4$ and $10^5$ years, these major reservoirs of carbon are gaseous C, CO, and \ce{C3}, which are clustered in the middle-right of Figures~\ref{fig:C-poor-Cfrac}-\ref{fig:C-rich-Cfrac}, and grain-mantle \ce{CH4}, \ce{H2CO}, \ce{HCN}, and \ce{CH3NH2}, which are clustered together in the upper-right. This is the same in both the oxygen-rich (Figure~\ref{fig:C-poor-Cfrac}) and oxygen-poor (Figure~\ref{fig:C-rich-Cfrac}) models, although there are small differences in time evolutions of these carbon fractions. In particular, the conversion of atomic carbon to \ce{C3} is significantly faster in the oxygen-poor model as evidenced by the more rapid disappearance of the dot corresponding to atomic carbon and more rapid growth of the dot corresponding to \ce{C3}. Starting from $10^5$ years, there are drastic and rapid changes to the form of carbon. The amount of carbon in atomic form and gas-phase \ce{C3} quickly drops, while the abundance of ice-phase CO grows as it freezes out from the gas phase. The carbon fractions of grain-mantle \ce{CH4}, \ce{HCN}, \ce{H2CO}, and \ce{CH3NH2} stay relatively constant between $10^5$ and $10^6$ years with some slight increases and decreases.

Aside from these changes the oxygen-rich and oxygen-poor models diverge significantly. In the oxygen-poor model, there are numerous dots in the bottom half of the figure that quickly grow in size over the range of $10^5$ to $\sim5\times10^5$ years. These represent various unsaturated carbon-chain molecules such as \ce{C8H2}, which constitutes slightly more than 2\% of the total carbon near the end of this range. Furthermore a sizeable fraction of total carbon is present in the two isomers of \ce{C3H4}, allene (\ce{CH2CCH2}) and propyne (\ce{CH3CCH}), in the grain mantle. \ce{CH2CCH2} can be seen as a major carbon reservoir even in the oxygen-rich model, however this becomes much more pronounced in in the oxygen-poor model. By $10^6$, years grain-mantle \ce{CH2CCH2} is the second largest reservoir of carbon at slightly more than 10\%. The hydrocarbons \ce{C2H6}, \ce{C3H8}, \ce{CH2CHCCH}, \ce{C6H4}, and \ce{C8H4} in the grain mantle additionally become significant carbon reservoirs in the oxygen-poor model as the simulation progresses to $10^7$ years. The model with a C/O ratio of 1 (Figure~\ref{fig:C-unity-Cfrac}) is an intermediate between the oxygen-rich and oxygen poor scenarios. Even in this model the faster conversion of atomic carbon to \ce{C3} and appearance of gas-phase carbon chains can be seen, although not to the degree of the oxygen-poor model. The time-dependent carbon reservoirs in our model are thus highly dependent on C/O ratio with gas-phase carbon chains and grain-mantle hydrocarbons becoming more abundant as the C/O ratio increases.

\newpage
\subsection{Fractions of carbon in the gas and grain phases and comparison with observations}
\label{sec:partitioning}

\begin{figure*}
    \centering
    \includegraphics[width=.3\textwidth]{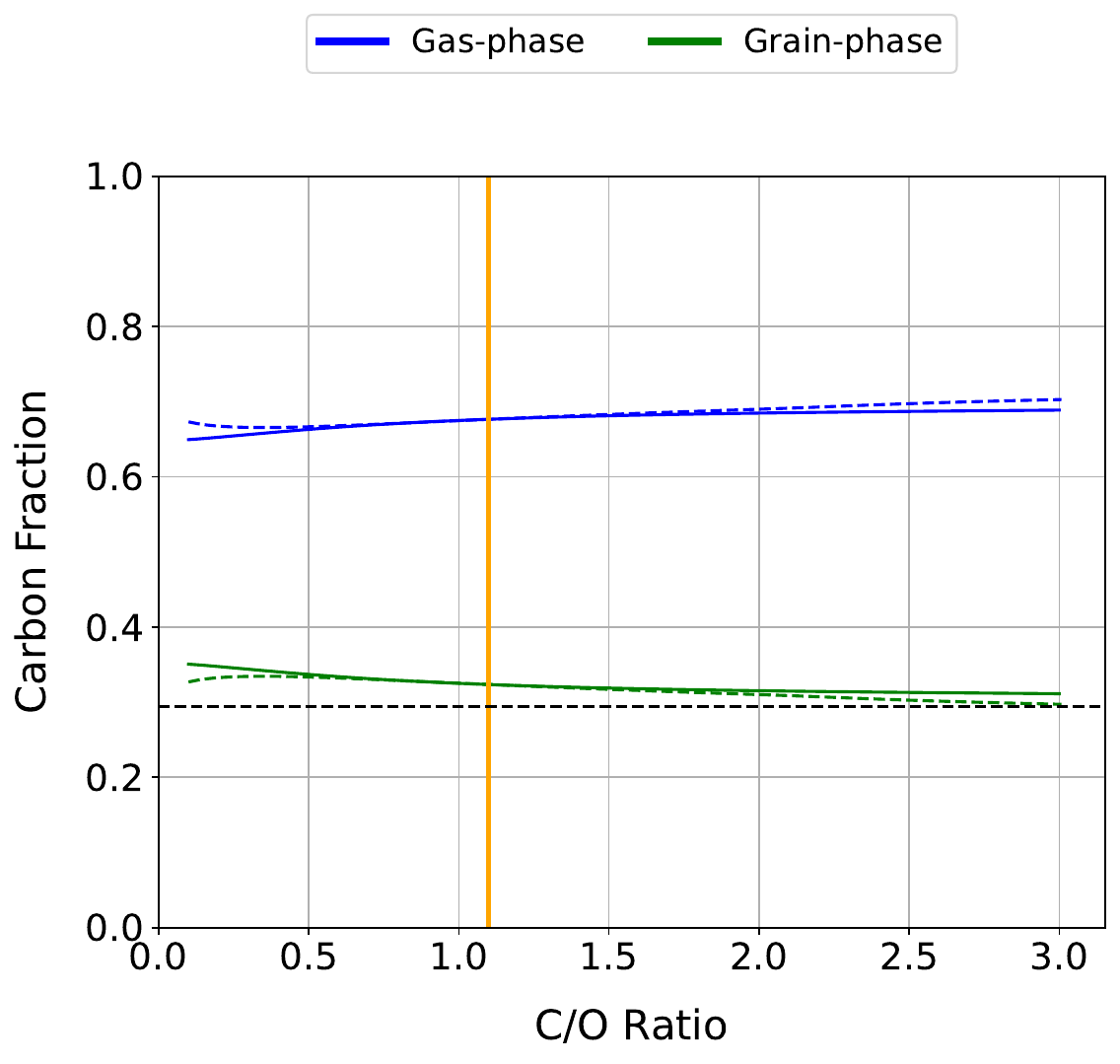}\hfill
    \includegraphics[width=.3\textwidth]{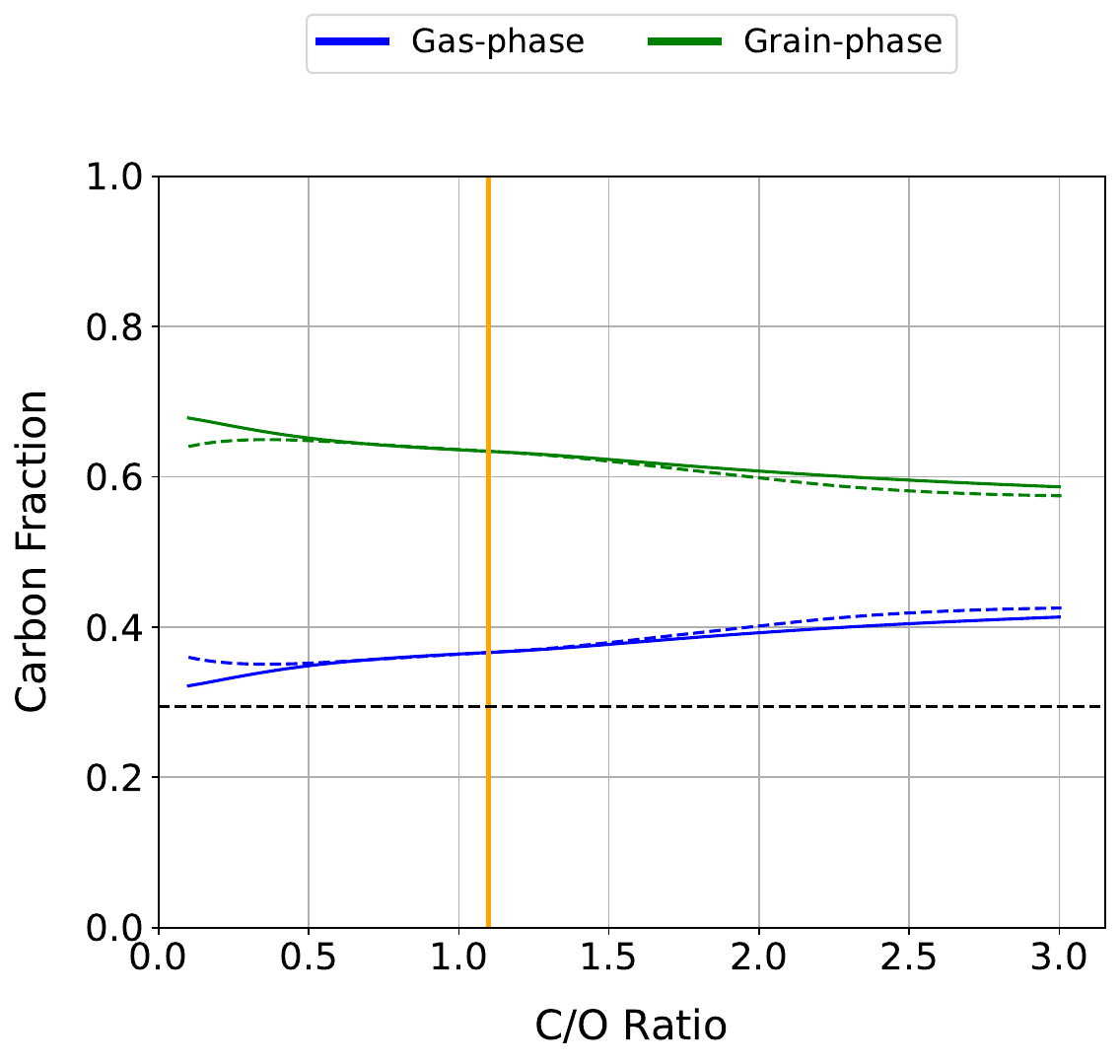}\hfill
    \includegraphics[width=.3\textwidth]{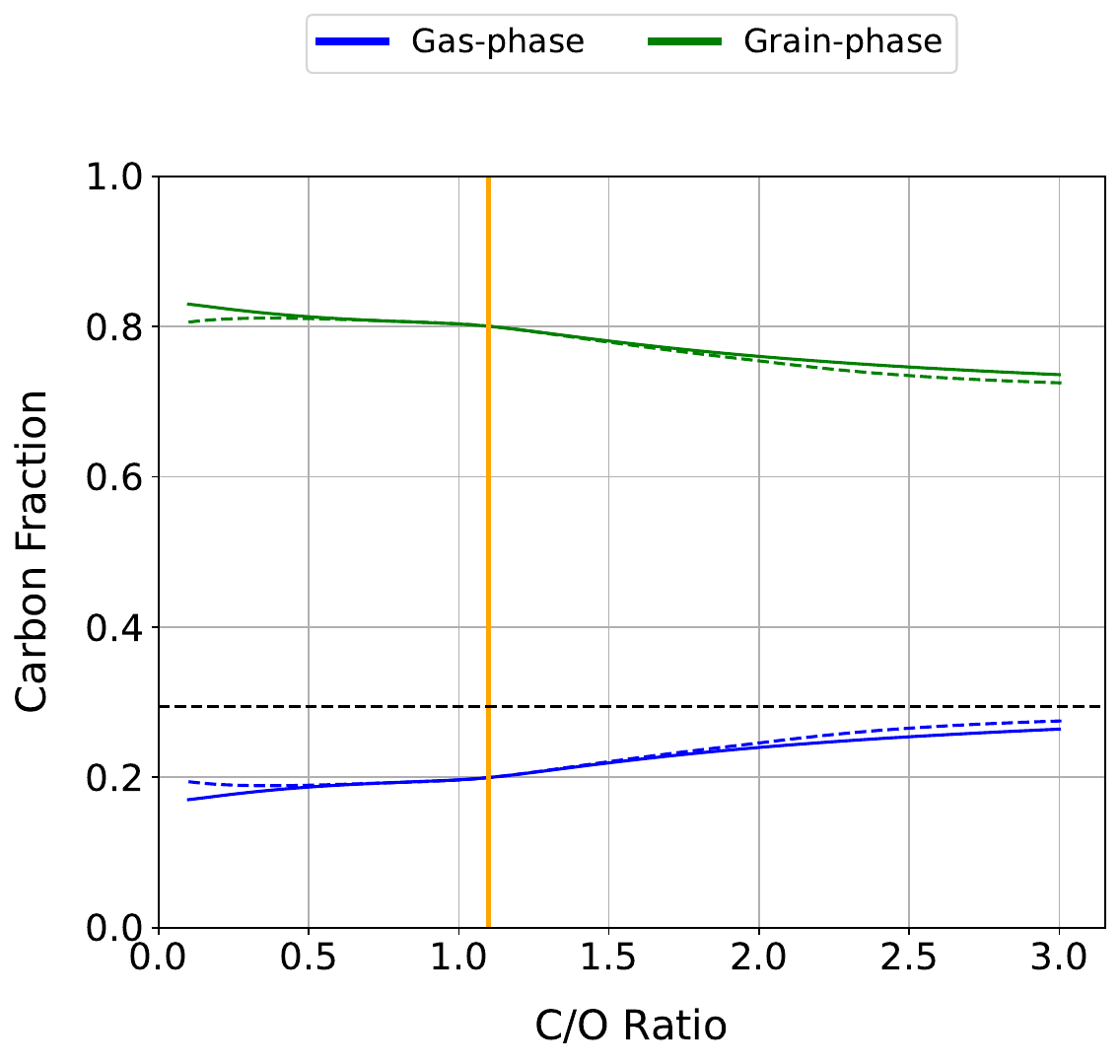}\hfill
    \caption{Modeled fractions of total carbon on the grains (green lines) and in the gas-phase (blue lines) at $1\times10^5$ (left), $3\times10^5$ (center), and $5\times10^5$ (right) years. The solid lines are from changing only the initial oxygen abundance, and the dashed lines are from changing only the initial carbon abundance. The black dashed line is the gas-phase carbon fraction based on the observed C/H from \citet{xue_molecular_2025} a total carbon abundance of $1.7\times10^{-4}$. A vertical orange line is plotted at our nominal C/O ratio of 1.1.}
    \label{fig:gas_grain_frac}
\end{figure*}

Recently \citet{xue_molecular_2025} have compiled and re-analyzed the column densities of many of the detected TMC-1 CP species to-date. The authors converted these column densities to abundances of carbon with respect to hydrogen nuclei, finding a total gas-phase carbon abundance of $5.0\times10^{-5}$ when summed together. Many chemical kinetic models of TMC-1 CP assume a total carbon budget of $1.7\times10^{-4}$ with respect to hydrogen nuclei based on observations of the diffuse cloud $\zeta$ Ophiuci \citep{jenkins_unified_2009, agundez_chemistry_2013} The observed value from \citet{xue_molecular_2025} is 29.4\% of this, suggesting that the detected, gas-phase molecular inventory of TMC-1 CP accounts for approximately 30\% of the total carbon budget if the $\zeta$ Ophiuci value is indeed representative. We note that this observation-based value does not consider uncertainties in the column densities. Although individual column densities are often well-constrained, there is a currently unquantified uncertainty in this gas-phase carbon budget due to the combination of these observational uncertainties. In Figure~\ref{fig:gas_grain_frac} we compare the observed gas-phase carbon fraction from \citet{xue_molecular_2025} with the gas- and ice-phase carbon fractions as obtained by our model. We plot the modeled fractions as a function of C/O ratio for both changing initial oxygen and carbon abundances and for simulation times of $1\times10^5$, $3\times10^5$, and $5\times10^5$. In each plot we show the value from \citet{xue_molecular_2025} as a horizontal black dashed line.

In all cases the gas-phase carbon fraction increases with increasing C/O ratio, although this is relatively small with a maximum difference of $\sim10\%$ over the entire range tested. At $1\times10^5$ years, approximately 70\% of the total carbon is in the gas-phase for all C/O ratios tested. The amount of gas-phase carbon quickly decreases as time goes on, dropping to 40\% at $3\times10^5$ years and only 20\% at $5\times10^5$ years. The observed value does not include every species that has been detected in TMC-1 CP. More importantly, it does not include a number of carbon-containing species that have not been detected but are known or thought to be present, such as \ce{CH3}, \ce{C3}, and even atomic carbon itself. These latter two species in particular are highly abundant forms of carbon that are included in our model along with a number of undetected carbon-containing species. We thus expect the observed value from \citet{xue_molecular_2025} to underestimate the gas-phase carbon fraction compared to the model. This is true at $1\times10^5$ and $3\times10^5$ where the modeled gas-phase carbon fraction is well above and slightly above the observed value respectively, but not at $5\times10^5$ years where the model is slightly below the observations.

\section{Discussion}
\label{sec:Discussion}
\subsection{Effects of C/O Ratio on Carbon Chemistry}

\begin{figure}
    \centering
    \includegraphics[width=\columnwidth]{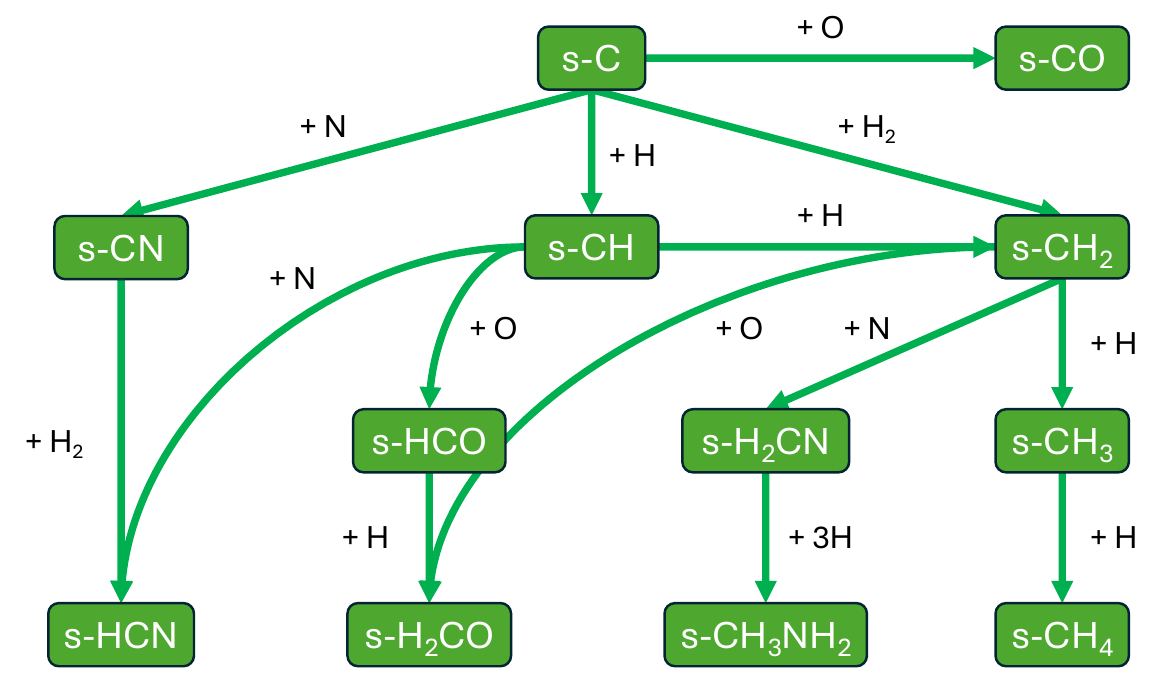}
    \caption{Chemical network leading to the formation of key ice-phase species in our model, beginning with atomic carbon. Each green arrow is a grain-surface reaction, with the other reactant labeled in black. The ``s-'' prefix denotes grain-surface species.}
    \label{fig:Cgrain_network}
\end{figure}

\begin{figure}
    \centering
    \includegraphics[width=\columnwidth]{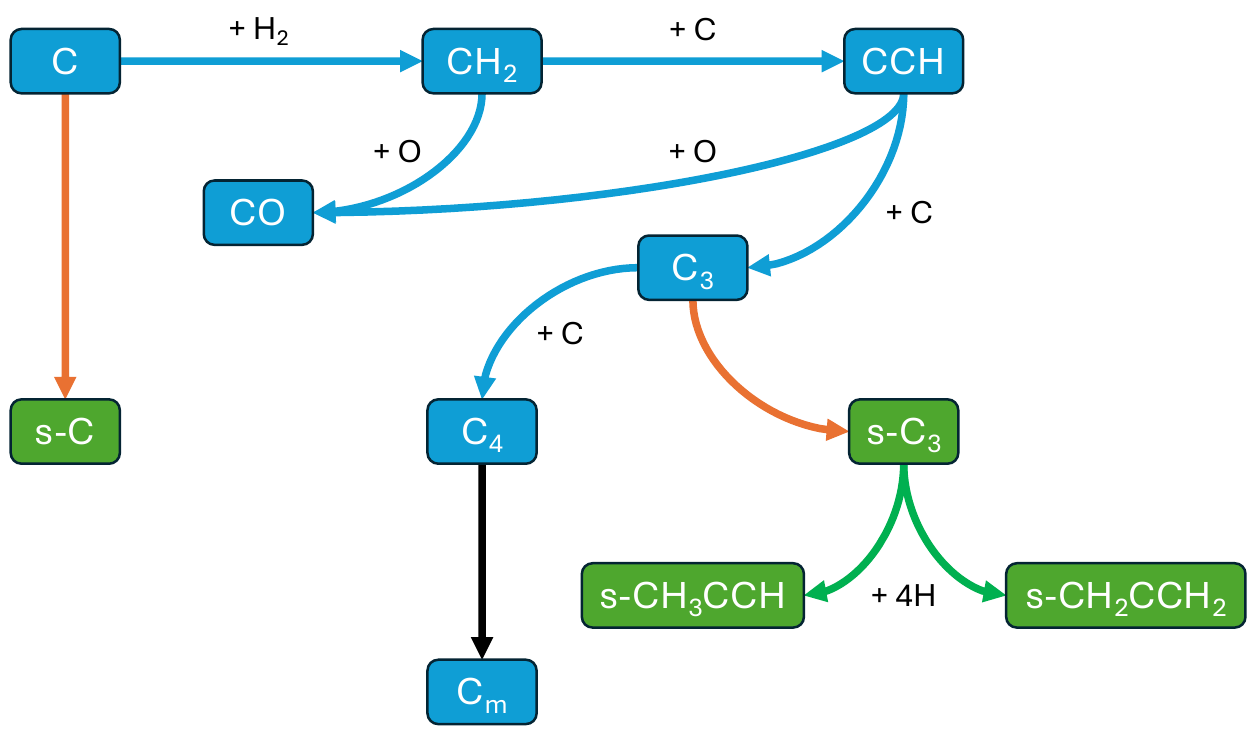}
    \caption{Chemical network leading to the formation of \ce{C3} and beyond in our model. Blue boxes and arrows denote gas-phase species and reactions respectively, while grain-surface species and reactions are in green. Grain-surface species are also labeled with the ``s-'' prefix. Orange arrows correspond to adsorption to grain surfaces. The black arrow represents carbon-chain growth according to Reactions~\ref{eqa:C+CmH} and \ref{eqa:Cm-+C}.}
    \label{fig:C3_network}
\end{figure}

As noted previously, the major carbon reservoirs are relatively insensitive across the range of astrophysically-relevant C/O ratios (between 0.5 and 2.0) and primarily consist of gas-phase C, \ce{CO}, and \ce{C3} as well as grain-mantle \ce{CO}, \ce{HCN}, \ce{CH3NH2}, \ce{CH4}, and \ce{H2CO}. Aside from CO, which can also be efficiently produced in the gas phase, these species are all formed predominately on the grain surface beginning with atomic carbon. The major formation pathways as identified based on modeled reactions fluxes can be seen in Figure~\ref{fig:Cgrain_network}. After adsorbing to a grain, atomic carbon can quickly react with other major grain-surface species. At early times these are primarily H, \ce{H2}, N, and O. An initial reaction with O or N leads to CO and CN respectively, the latter of which undergoes hydrogenation to HCN via \ce{H2}. This is consistent with \citet{ruaud_gas_2016}, who likewise found a large abundance of ice-phase \ce{HCN} in their gas-grain model due to the \ce{H2 + CN} reaction when competition between reaction and diffusion is considered. HCN can also be efficiently formed in the gas phase, although its adsorption to grains competes with faster ion-neutral destruction mechanisms \citep{loison_interstellar_2014}. Successive hydrogenation of atomic carbon leads to \ce{CH4}, although reaction of CH and \ce{CH2} with N and O are competing processes. Addition of O either forms \ce{HCO}, which is hydrogenated to \ce{H2CO}, or \ce{H2CO} directly. Addition of N to CH and \ce{CH2} form HCN and \ce{H2CN} respectively. The latter is then fully hydrogenated up to \ce{CH3NH2} by successive additions of atomic hydrogen. Reactions involving atomic hydrogen are generally the fastest grain-surface processes due to its high mobility; however, reactions with O and N are competitive in our model because of their low assumed binding energies of 800 K, compared to 650 K for H. This 800 K value originates from \citet{tielens_model_1982} where it was estimated for C, N, and O. The binding energy of C has since been updated to 4000 K in the model we use \citep{ruaud_modelling_2015}, whereas the binding energies of N and O remain at 800 K. More recent computational and experimental work has indicated that the binding energy of O is closer to 1600 K \citep{he_new_2015, minissale_direct_2016, wakelam_binding_2017, shimonishi_adsorption_2018, minissale_thermal_2022}, although the binding energy of N has been found to be near or even lower than this original estimate \citep{minissale_direct_2016, wakelam_binding_2017, shimonishi_adsorption_2018, minissale_thermal_2022}. It has also been suggested more recently that C undergoes chemisorption rather than physisorption, resulting in even higher binding energies of 10,000+ K \citep{wakelam_binding_2017, shimonishi_adsorption_2018, minissale_thermal_2022}. Due to the difficulty in many astrochemical kinetic models assume that the diffusion barriers are related to the binding energies by some universal, constant factor. At a C/O ratio of 2, grain-surface hydrogenation of atomic carbon to \ce{CH4} becomes more efficient because there is less competition from reactions with atomic nitrogen and oxygen. This is likely because the increased amount of reactive, gas-phase carbon at larger C/O ratios leads to more participation of nitrogen and oxygen in gas-phase chemistry at early times and less adsorption onto grain surfaces. 

Although these major reservoirs of carbon do not change significantly with varying C/O ratio, there are a number of other noticeable changes. The faster disappearance of atomic carbon as it is converted to carbon-containing molecules, increased abundance of \ce{C3}, and decreased abundance of CO with increasing C/O ratio are all signs of enhanced carbon chemistry. This is in line with previous studies \citep{leung_synthesis_1984, blake_molecular_1987, wakelam_sensitivity_2009, hincelin_oxygen_2011, agundez_chemistry_2013}, as more available carbon relative to oxygen results in more carbon that is not ``locked up'' in CO and thus able to participate in carbon chemistry. The most striking change is that at C/O ratios above unity, a large number of unsaturated carbon-chain molecules quickly become sizable reservoirs of carbon. These carbon-chain species are sequentially built up via additions of atomic carbon, beginning with the reaction:
\begin{equation}
    \label{eqa:O+C4}
    \ce{C + C3 -> C4 + h$\nu$}.
\end{equation}
Although this is a radiative association mechanism between two small species, previous Rice-Ramsperger-Kessel-Markus (RRKM) calculations suggest a rate coefficient on the order of $10^{-12}$ at 10 K \citep{wakelam_sensitivity_2009}. Sensitivity analyses of TMC-1 CP kinetic models confirm that this reaction plays a significant role in the network \citep{wakelam_sensitivity_2009, byrne_sensitivity_2024}. As shown in Figure~\ref{fig:C3_network} based on modeled reaction fluxes, formation of \ce{C3} requires two additions of atomic carbon and competes with addition of atomic oxygen to form CO. Formation of larger carbon chains occurs primarily through the reactions
\begin{align}
    \ce{C + C_{m}H &-> C_{m+1} + H} 
    \label{eqa:C+CmH} \\
    \ce{C + C_{m}- &-> C_{m+1} + e-}.
    \label{eqa:Cm-+C}
\end{align}
In addition to requiring repeated additions of atomic carbon, all \ce{C_{m}H} and \ce{C_{m}-} species are quickly destroyed via reaction with atomic oxygen in the model, along with \ce{C_{m}} species with even-numbered $m$ \citep{loison_gas-phase_2014}. It therefore stands to reason that at low C/O ratios, where there is little carbon available relative to oxygen, these mechanisms will be inefficient. Conversely, the sudden appearance of various carbon-chain species at C/O ratios above one shows that these pathways are considerably important when the amount of available carbon outweighs that of oxygen. This is further confirmed by looking at the contributions of all reactions to the destruction of these species. At a C/O ratio of 0.5, this mechanism is the first or second most influential destruction reaction and accounts of $\sim20-40\%$ of the destruction. When the initial oxygen abundance is depleted to a C/O ratio of 2, it instead accounts for $5\%$ or less of the destruction at later times and is outpaced by reactions with atomic nitrogen, atomic carbon, and small ions such as \ce{H3+} and \ce{HCO+}.

The identification of ice-phase allene (\ce{CH2CCH2}) and propyne (\ce{CH3CCH}) as species that contain a significant fraction of total carbon, even at low C/O ratios, is surprising. These two \ce{C3H4} isomers are pure hydrocarbons, the former of which has no permanent dipole moment and the latter of which has been detected in multiple interstellar regions including TMC-1 CP\citep{kuiper_methyl_1984}. Previous modeling work centered on these species has focused on their gas-phase chemistry \citep{byrne_astrochemical_2023} but did account for grain-surface formation via sequential hydrogenation of \ce{C3} as indicated by \citet{hickson_methylacetylene_2016} and \citet{loison_interstellar_2017}. These first four additions of atomic hydrogen are barrierless, however the subsequent hydrogenations to \ce{C3H5} are expected to have a sizable activation barriers based on the corresponding gas-phase reactions. These barriers in conjunction with efficient formation of gas-phase \ce{C3} even at low C/O ratios are thus likely why ice-phase \ce{CH2CCH2} and \ce{CH3CCH} are able to build up in large abundance in the model. Between the two, the amount of carbon present in \ce{CH2CCH2} is two or more times that of \ce{CH3CCH}, reaching a staggering 10\% of total carbon at large C/O ratios. This is likely due to incomplete inclusion of \ce{CH2CCH2} into the chemical network. There are numerous ice-phase reactions of \ce{CH3CCH} that are missing for \ce{CH2CCH2}, including hydrogenation to \ce{C3H5} and beyond. Even so, the presence of these species as major carbon reservoirs is further evidence to the important of \ce{C3H$_n$} hydrocarbons in cold cloud chemistry, for which a major formation route is grain-surface hydrogenation starting from \ce{C3}. 

Despite being the most affected molecules when the C/O ratio is varied from 0.5 to 2, naphthalene and related species are not major reservoirs of carbon. Even with a C/O ratio of 2, the maximum carbon fraction obtained by naphthalene is $\sim1.5\times10^{-4}$\%. This is far below the estimated carbon fraction of 0.04\% for pyrene and coronene based on observations of the corresponding CN-substituted species \citep{wenzel_detection_2024, wenzel_detections_2024, wenzel_discovery_2025}. Although increasing the C/O ratio does provide nonlinear increases to the abundance of \ce{C10H8}, it alone is not enough to reproduce the abundance of PAHs that are observed in TMC-1 CP. Thus there are important chemical processes that are unaccounted for in the model that convert a significant amount of carbon into these species. In particular, \ce{C3H$_n$} hydrocarbons are precursors to more complex species like benzene (\ce{C6H6}) and naphthalene (\ce{C10H8}) \citep{byrne_astrochemical_2023}. These species can clearly form in large abundance on grain surfaces in the model. Future work should investigate additional grain-surface reactivity of \ce{C3H$_n$} species as well as their desorption mechanisms. 

\subsection{Partitioning of Carbon into Gas and Grain Phases}
As mentioned in Section~\ref{sec:partitioning}, the approximate gas-phase C/H ratio of TMC-1 CP based on observations cannot be reproduced by the model at $5\times10^5$ years, even under extreme oxygen-poor conditions. The rapid decrease in the modeled gas-phase carbon fraction over this period of time is due to the freeze out of gaseous molecules, primarily CO, onto grain surfaces. The much larger gas-phase carbon fractions at earlier times of $1\times10^5$ and $3\times10^5$ years could suggest that there has been less time for freeze out of gas-phase molecules onto interstellar dust grains in clouds like TMC-1 CP, resulting in lower chemical ages. Additionally, the modeled abundance of gaseous CO, the most abundant carbon-containing molecule, agrees best with observations at an ``intermediate'' time of $3\times10^5$ years. However, it has been shown that the abundances of numerous complex organic molecules such as cyanopolyynes and aromatic species cannot be reproduced until later times closer to $5\times10^5$ years \citep{loomis_investigation_2021, byrne_astrochemical_2023}. Thus there is a discrepancy between the times at which a substantial fraction of carbon is in the gas-phase and the times at which the modeled kinetics can reach a greater degree of chemical complexity. 

One possible solution is that the freeze-out of carbon-containing molecules from the gas phase onto grain surfaces is overly efficient in the model. As shown by the large carbon fractions of ice-phase \ce{CH2CCH2} and \ce{CH3CCH}, large abundances of carbon-containing molecules, including precursors to PAHs and other COMs, can build up on interstellar grains by later times in the model. Grain-surface reactions in the chemical network primarily consist of hydrogenation via hydrogen atoms due to its fast diffusion at 10\,K. Carbon atoms have also been shown experimentally to quickly diffuse on low temperature amorphous solid water, although this does not occur until 22\,K and above \citep{tsuge_surface_2023}. For grain-surface processes to contribute to the gas-phase chemical inventory, there must be efficient desorption. Thermal evaporation and photodesorption via external UV photons are slow in molecular clouds due to the low temperatures and large densities. Consequently photodesorption via the internal UV field, reactive desorption due to reaction exothermicity, and occasionally cosmic-ray heating of grains are the major processes, with the efficiency of reactive desorption assumed to be 1\% \citep{garrod_non-thermal_2007}. This 1\% efficiency is the lower-end of the estimated $0.01 - 0.1$ range often used in kinetic models. In actuality, this parameter varies from reaction to reaction and is difficult to quantify experimentally \citep{minissale_dust_2016, chuang_reactive_2018}. Desorption via cosmic-ray sputtering and grain-grain collisions have also been tested in astrochemical models \citep{wakelam_efficiency_2021, kalvans_icy_2022, wakelam_2024_2024, rosandi_astrophysical_2024}. These mechanisms are both capable of increasing modeled gas-phase abundances of COMs by multiple orders of magnitude under dense conditions. Further benchmarking of these processes in kinetic models with complex chemistry is necessary.

Another possibility is that there are additional sources of reactive, gas-phase carbon that are not presently considered in the model. As shown in the rightmost panel of Figure~\ref{fig:gas_grain_frac}, we find that increasing the C/O ratio up to 3.0 by increasing the initial carbon abundance still cannot reach a gas-phase carbon fraction of 30\%. However, in these scenarios the total carbon budget is greater, and thus the absolute amount of carbon in the gas phase is able to increase above the value determined by \citep{xue_molecular_2025}. Such an increase could potentially be achieved by destruction of carbonaceous grains or even large PAHs if the top-down mechanism is operative, although a $\sim1.36x$ increase to the total carbon budget is required at minimum. In a recent sensitivity analysis \citep{byrne_sensitivity_2024}, we found that many carbon-containing molecules in the model are highly sensitive to the rate of reaction between atomic nitrogen and CN. This is likely because this process frees up atomic carbon from CN that is subsequently used up to form more complex carbon molecules. More processes like this could be responsible for transforming relatively inert forms of carbon species to reactive building blocks, possibly shortening the timescale of formation for PAHs and other COMs. In particular the main reservoir of carbon is in CO; however the only major processes known to break apart CO in molecular clouds are cosmic ray dissociation and ion-neutral reaction with \ce{He+}. These explanations thus seem less plausible than uncertainties in grain-surface processes.

\subsection{Astrochemical Implications}

\begin{figure}
    \centering
    \includegraphics[width=\columnwidth]{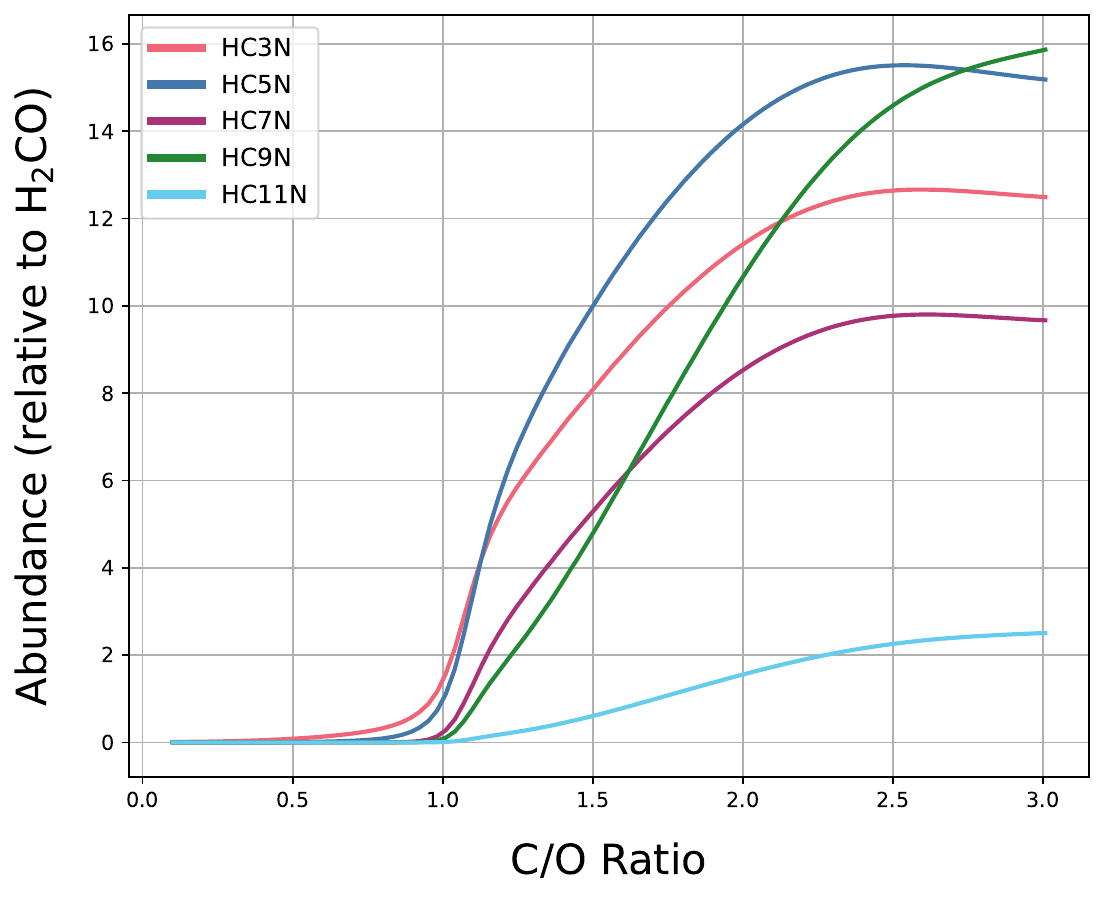}
    \caption{Abundance ratios of detected cyanopolyyne species to \ce{H2CO} at $5\times10^5$ years as a function of C/O ratio. Each color represents a different cyanopolyyne.}
    \label{fig:cyanopolyynes_norm}
\end{figure}

According to the sensitivities shown in Figure~\ref{fig:C_O_Sens}, unsaturated, carbon-rich molecules such as polyynes and PAHs are some of the most sensitive species to the modeled C/O ratio. Similarly, \citet{burkhardt_ubiquitous_2021} compared observed abundances of \ce{HC9N} and \ce{C6H5CN} across various sources, finding that the abundance ratios of \ce{HC7N} and \ce{HC9N} to benzonitrile are up to five times larger in TMC-1 CP compared to sources in the Serpens cloud. Using astrochemical modeling the authors also showed that the cyanopolyynes more drastically decrease in abundance as the C/O ratio is lowered compared to \ce{C6H5CN}. The large abundance of cyanopolyynes relative to \ce{C6H5CN} could thus be evidence of TMC-1 CP having a C/O ratio above unity. It has previously been suggested that abundance gradients within the Taurus Molecular Cloud are due to a variation in C/O ratio, with a greater C/O ratio at the cyanopolyyne peak \citep{pratap_study_1997}. Our findings are also consistent with the idea that a C/O ratio above 1 better reproduces the abundances of complex hydrocarbons observed toward this position \citep{wakelam_effect_2006, loomis_investigation_2021, burkhardt_ubiquitous_2021}. However, other recent modeling studies have indicated that a lower C/O ratio of 0.5-0.7 can provide better agreement with observations \citep{agundez_chemistry_2013, loison_gas-phase_2014, chen_chemical_2022}. \citet{agundez_chemistry_2013} and \citet{chen_chemical_2022} both found that a diffuse cloud C/O ratio of $\sim0.5$ best reproduces observed abundances in TMC-1 CP. The authors of both studies attribute this to the overproduction of large carbon chains, such as \ce{C$_n$H-} and \ce{C$_n$H2} species, between $10^5$ and $10^6$ years when the C/O ratio is above one. As mentioned by \citet{chen_chemical_2022}, this may be due to a lack of sufficient destruction mechanisms for these large, unsaturated hydrocarbons. However, their best-fit model of uses an elemental carbon abundance of $2.5\times10^4$ with respect to H nuclei, which is larger than the value of $\sim1.7\times10^4$ typically used. Additionally, the chemical networks used in these studies did not include recently detected COMs such as aromatic species that are highly sensitive to the C/O ratio. PAHs have been proposed to be formed through two major scenarios: ``bottom-up'' chemistry forming these molecules from smaller precursors, or ``top-down'' destruction of larger PAHs and carbonaceous dust such as through photodissociation by UV photons. Our model assumes that PAHs are formed solely through bottom-up processes and does not consider any top-down formation. However, the relatively flat relationship between size and column density for PAHs detected in TMC-1 CP \citep{wenzel_discovery_2025} may indicate that top-down processes are operative. If so, relationship between the C/O ratio and PAH abundances could be significantly different than that presented here, although the contributions of these two scenarios in dense clouds are still unclear.

It is also possible that information regarding the elemental abundances of carbon and oxygen could be extracted by comparison to abundances that are expected to be insensitive to the C/O ratio. \ce{H2CO} is the least-sensitive carbon-bearing molecule to the C/O ratio that has been detected in the ISM, where it has been observed toward many objects \citep{snyder_microwave_1969, zuckerman_observations_1970, dieter_survey_1973, mangum_formaldehyde_2008}. Although gas-phase formation is more efficient at lower C/O ratios due to the greater availability of oxygen atoms, grain-surface formation via hydrogenation of CO becomes faster at larger C/O ratios due to an increased abundance of CO on grain surfaces. Conversely, cyanopolyynes have also been widely observed \citep{turner_detection_1971, chapillon_chemistry_2012, taniguchi_survey_2018, mendoza_search_2018, loomis_investigation_2021} but are highly sensitive to the C/O ratio in kinetic models. In Figure~\ref{fig:cyanopolyynes_norm} the modeled abundances of cyanopolyynes up to \ce{HC11N} relative to the modeled abundance of \ce{H2CO} are shown as a function of C/O ratio at the time of best fit. Above a C/O ratio of 1, all cyanopolyynes except \ce{HC11N} become more abundant than \ce{H2CO} with a rapidly increasing abundance ratio. Below a C/O ratio of 1 \ce{H2CO} is more abundant, although \ce{HC3N} is abundant enough that this ratio is significant and highly influenced by a change to C/O ratio. Given the ubiquity of these species throughout the interstellar medium, the abundance ratios of cyanopolyynes, particularly \ce{HC3N}, to \ce{H2CO} might be strong probes of the C/O ratios of astronomical sources.

In terms of major carbon reservoirs, inheritance disk models often consider CO, \ce{CO2}, \ce{CH4}, and occasionally \ce{CH3OH} and \ce{HCN} \citep{eistrup_setting_2016}. This is consistent with previous kinetic models of interstellar ices, in which most of the ice-phase carbon ends up as CO, \ce{CO2}, \ce{CH4}, and \ce{CH3OH} \citep{garrod_formation_2011, aikawa_chemical_2020}. Likewise, recent JWST observations of molecular cloud ices have found these four molecules to be the most abundant ice-phase carbon reservoirs, although some evidence of complex organic molecules is also seen \citep{yang_corinos_2022, mcclure_ice_2023}. Our model agrees that CO and \ce{CH4} constitute major carbon reservoirs on interstellar grains. This occurs under all three C/O ratios of 0.5, 1.0, and 2.0, where these species each contain 2\% of the total carbon by $\sim5\times10^5$ years and grow in abundance as time progresses until the end of the simulation ($10^7$ years). However, \ce{CO2} and \ce{CH3OH} do not appear as major ice-phase species. Ice-phase \ce{CO2} primarily forms through reaction of CO with OH radicals, which does not become efficient until 12 K and warmer when CO becomes mobile unless non-diffusive mechanisms are invoked \citep{garrod_formation_2011}. As our simulation uses a grain temperature of 10 K and does not consider non-diffusive mechanisms, \ce{CO2} is unable to be efficiently formed. Likewise, \ce{CH3OH} is typically formed from hydrogenation of CO through successive additions of H atoms; however the first and third steps to form HCO and \ce{CH3O} respectively have been found to have sizable energy barriers \citep{fuchs_hydrogenation_2009}. Furthermore, H-atom abstraction to reform HCO is a competing pathway \citep{hidaka_reaction_2009, chuang_h-atom_2016} that uses a lower barrier than the addition channels in our model \citep{ruaud_modelling_2015}. Conversely, our model does suggest that complex hydrocarbons like \ce{C3H4}, \ce{C3H8}, and \ce{CH2CHC2H} can be abundant in ices under oxygen-poor conditions. More work is needed to build a robust, ice chemistry network that can investigate the interplay between formation of simple and complex organic molecules.

In protoplanetary disks, the C/O ratio of the disk and its resulting planets have a strong spatial dependence \citep{oberg_effects_2011}. Detections of complex hydrocarbons such as \ce{C2H6}, \ce{C3H4}, and even benzene in inner disks with JWST are challenging the previous, canonical C/O ratio of 0.45. As with molecular clouds like TMC-1 CP, hydrocarbon abundances in chemical kinetic disk models are highly sensitive to the C/O ratio \citep{oberg_effects_2011, du_volatile_2015, kanwar_minds_2024}, with a large C/O ratio often necessary to reproduce the observed hydrocarbon abundances \citep{bergin_hydrocarbon_2016, kanwar_minds_2024}. Additionally, the disk C/O ratio can strongly influence the physics, such as planetesimal formation \citep{xenos_how_2025}. It is not known whether the initial chemical conditions, such as the presence of molecules and/or an enhanced gas-phase C/O ratio, are inherited from the molecular cloud or if there is a reset back to diffuse cloud, atomic elemental abundances \citep{eistrup_setting_2016, eistrup_molecular_2018, pacetti_chemical_2022}. 

It is possible that such carbon-rich conditions were already present in the previous molecular cloud or that the mechanisms for enhancing carbon relative to oxygen are similar between the cloud and the disk. Since the solar abundance of oxygen is almost two times that of carbon \citep{lodders_solar_2003}, a C/O ratio above unity requires some kind of mechanism that more strongly depletes oxygen. In \citet{furuya_water_2015}, the authors simulate the formation of a molecular cloud to investigate the early-stage water formation. They find that water ice can form quickly even at low visual extinction due to grain-surface hydrogenation of atomic oxygen. As the visual extinction increases and the temperature decreases, desorption of water ice becomes less efficient and the gas-phase abundance of elemental oxygen drops significantly. The formation of gas-phase CO happens on approximately the same timescale. As pointed out by \citet{sellek_chemical_2025}, liberation of carbon from CO is also necessary to obtain a C/O ratio above unity. This could be done via reaction with \ce{He+} as they describe and investigate. Alternatively, it is possible that carbon could be liberated from the processing of carbonaceous grains. For example, \citet{bocchio_re-evaluation_2014} calculated the dust lifetime of carbonaceous and silicate grains exposed to interstellar shocks, finding that carbonaceous grains have a shorter lifetime. Another possibility proposed by \citet{blake_molecular_1987} is hydrogenation of both C and O on grain surfaces, with the resulting \ce{CH4} being able to desorb much more efficiently than \ce{H2O}. Recent work from the GEMS survey studied the relationship between elemental depletion and visual extinction in molecular clouds using observations of molecular emission and chemical modeling \citep{fuente_gas_2019}. The authors found that within the translucent phase, the gas-phase abundances of carbon and oxygen decrease with visual extinction as they combine to form CO and then CO freezes out onto grains. They also find that a C/O ratio of approximately unity describes their observations of the translucent phase well, requiring an oxygen depletion larger than the ``maximum'' depletion case of \citet{jenkins_unified_2009}. This seemingly corroborates the idea of a greater oxygen depletion in TMC-1 CP, although it is unclear whether the C/O ratio can be further increased above unity.


\section{Conclusions}
\label{sec:Conclusions}
We explored the effects of varying C/O ratio ranging from 0.1 to 3.0 on modeled hydrocarbon chemistry using the \texttt{NAUTILUS} 3-phase model and a large network containing many complex hydrocarbons, including the PAH naphthalene. Sensitivities of modeled abundances to C/O ratio were determined using relative standard deviation. These depend on the C/O range considered, with unsaturated carbon-chains and bicyclic aromatic molecules being the most affected under astrophysically-relevant conditions. Despite this sensitivity, PAH abundances cannot be reproduced by increasing the C/O ratio. The major carbon reservoirs were visualized using machine learning embedding and dimensionality reduction techniques. These instead consist of simple grain species including \ce{CO}, \ce{CH4}, \ce{H2CO}, \ce{HCN}, and \ce{CH3NH2} at both low and high C/O ratios. At C/O ratios above unity, there is a significant enhancement of complex hydrocarbon chemistry and the formation of unsaturated carbon-chains. The large abundances of \ce{C3H$_n$} hydrocarbons demonstrate the importance of these molecules, although the network surrounding these species is not complete. Furthermore, the model struggles to reproduce the gas-phase fraction of carbon based on observations of TMC-1 CP. More work regarding the chemistry of \ce{C3H$_n$} and the role they play in PAH formation, as well as general desorption mechanisms and sources of reactive carbon, is needed. Further investigation is also required to understand the connection between the chemical conditions and inventory of molecular clouds and planet-forming disks.

\section{Acknowledgements}
We thank Dr. Valentine Wakelam for use of the \texttt{NAUTILUS} v1.1 code. We also thank Dr. Hannah Toru Shay, Dr. Aravindh N. Marimuthu, and Zachary T. P. Fried for helpful discussions regarding the machine learning pipeline as well as Dr. Haley N. Scolati and Dr. Kin Long Kelvin Lee for the pipeline itself. A.N.B. acknowledges support from the National Science Foundation Grant Number 2141064. G.W., M.S.H., and B.A.M. acknowledge support from an Arnold and Mabel Beckman Foundation Beckman Young Investigator Award and from the Schmidt Family Futures Foundation.  B.A.M. and C.X. acknowledge support of National Science Foundation grant AST-2205126. The National Radio Astronomy Observatory is a facility of the National Science Foundation operated under cooperative agreement by Associated Universities, Inc. The Green Bank Observatory is a facility of the National Science Foundation operated under cooperative agreement by Associated Universities, Inc. 

\clearpage

\bibliographystyle{aasjournal}
\bibliography{main}

\end{document}